# Tracking the Mn diffusion in the carbon-supported nanoparticles through the collaborative analysis of atom probe and evaporation simulation

Chanwon Jung [a,b,†], Hosun Jun [a,†], Kyuseon Jang [a], Se-Ho Kim [b], Pyuck-Pa Choi [a,*]

[a] Department of Materials Science and Engineering, Korea Advanced Institute of Science and Technology (KAIST), 291 Daehak-ro, Yuseong-gu, Daejeon 34141, Republic of Korea

[b] Max-Planck-Institut für Eisenforschung GmbH, Düsseldorf 40237, Germany

[†] These authors contributed equally to this work

[*] Corresponding author: p.choi@kaist.ac.kr

**Key words**

Atom probe tomography, local magnification effect, carbon-supported nanoparticles, field evaporation simulation, intermetallic compounds

# Abstract

Carbon-supported nanoparticles have been used widely as efficient catalysts due to their enhanced surface-to-volume ratio. To investigate their structure-property relationships, acquiring 3D elemental distribution is highly required. Here, the carbon-supported Pt, PtMn alloy, and ordered $Pt_3Mn$ nanoparticles are synthesized and analyzed with atom probe tomography as model systems. The significant difference of Mn distribution after the heat-treatment was found. Finally, the field evaporation behavior of the carbon support was discussed and each acquired reconstruction was compared with computational results from the evaporation simulation. This paper provides a guideline for studies using atom probe tomography on the heterogeneous carbon-nanoparticle system that leads to insights toward to a wide application in carbon-supported nano-catalysts.

# Introduction

Atom probe tomography (APT) is a powerful analytical method for mapping the three dimensional (3D) atomic distribution and determining the local chemical composition of materials with sub-nanometer spatial resolution and ppm-level elemental sensitivity (Kelly & Larson, 2012; Miller, 2012; Kelly & Miller, 2007; Gault et al., 2012). Due to its unique capability, this technique has been adopted to reveal grain boundary segregation (Hojin Lee et al., 2019; Kim et al., 2019), nano-precipitation (Kürnsteiner et al., 2017), and chemical compositions of nano-structures for various bulk materials (Gault et al., 2018; Jung et al., 2019). Since investigation of chemical composition and information of 3D elemental distribution is highly important to understand property-structure correlation in nanoparticles, there have been many attempts to analyze using APT (Barroo et al., 2019; Changsoo Lee et al., 2019). Tedsree et al. analyzed Ag-Pd core-shell metal nanoparticles to reveal the relationship between formic acid decomposition and Pd shell thickness (Tedsree et al., 2011). Li et al. performed APT measurements on the carbon-supported Pt, Pt/Co alloy, and Ir-Pt core-shell nanoparticles by depositing the nanoparticles to pre-sharpened tips using electrophoresis (Li et al., 2014). Felfer et al. used PVD-Cr sputtering on the Au-Ag core-shell nanoparticles dispersed on the Si substrate to embed the nanoparticles (Felfer et al., 2014). More recently, Kim et al. developed the co-electrodeposition of a metal to embed various nanomaterials including freestanding hollow $TiO_2$ nanowire (Lim et al., 2020), $MoS_2$ nanosheet (Se-Ho Kim et al., 2020), $Li_4Ti_5O_{12}$ (Kim et al., 2022), Pd nanoparticles (Kim et al., 2018; Jang et al., 2021), and Pd@Au core@shell nanoparticles (Se-Ho Kim et al., 2020).

While successful APT measurements of the nanoparticles have been performed using abovementioned methods, the discussion on the analyzed data are mainly focused on nanoparticles, and the effect of carbon support has often been neglected. The community has recently started to expand the analysis capabilities to carbon-based materials such as carbon fiber (Johansen & Liu, 2021; Johansen et al., 2021), carbon nanotube (Raghuwanshi et al., 2020), diamond (Schirhagl et al., 2015), graphene (Exertier et al., 2021) and self-assembled decanethiol molecules (Gault et al., 2010; Stoffers et al., 2012; Solodenko et al., 2021) not limited to metals or intermetallic compounds. However, since elemental carbon has the highest intrinsic evaporation field compared to any metallic element, the accurate data interpretation of carbon materials *i.e.* the carbon black support is challenging; the evaporation field value of covalent-bonded carbon calculated to be 103 V nm$^{-1}$ whereas the most metals shows theoretical evaporation fields in the range of 15 to 60 V nm$^{-1}$ (Larson et al., 2013; Gault et al., 2012). As a result, the presence of a carbon support adjacent to a metallic nanoparticle will generate inhomogeneous evaporation behavior at the specimen's surface and eventually result a distorted 3D atom map from the actual specimen, which is known as local magnification effect (LME). LME is critical to the data reconstruction algorithm since it influences atomic density fluctuation, interface shape distortion, and apparent intermixing of phases in acquired data that may provide mis-interpretation. The area of the LME affected region could reach up to scale of few nanometers (Madaan et al., 2015) depending on both composition and structure, thus, it is crucial to study the artifacts caused by carbon support on the nanoparticles and validate the reliability of the carbon-nanoparticle APT data.

Hence, for demonstration, we have chosen a series of Mn/Pt nanoparticles that are known as effective oxygen reduction electro-catalysts (Peng et al., 2020) as our model systems for investigation of the carbon-nanoparticle field evaporation. We compared the acquired APT data of each nanoparticle sample with numerical simulation using TAPSim software developed by Oberdorfer et al. (Oberdorfer et al., 2013) systematically. With a better understanding of the severe aberration of carbon, the Mn and Pt elemental distributions are subsequently correctly interpreted.

# Experimental section

## Materials

Manganese (III) acetylacetonate (Mn(acac)$_3$, technical grade), 2-hexadecanediol (HDD, 90 %), and diphenyl ether (99 %) were purchased from Aldrich. Cobalt sulfate heptahydrate (CoSO$_4$·7H$_2$O, 98 %), cobalt chloride hexahydrate (CoCl$_2$·6H$_2$O, 97 %), and boric acid (H$_3$BO$_3$, 99.5 %) were purchased from Daejung Chmicals & Metals. All other reagents were purchased from Samchun Chemical. All chemicals were used as received without further purification.

## Preparation of carbon-supported nanoparticles

To synthesize PtMn/C catalysts, 12 mg of manganese (III) acetylacetonate (Mn(acac)$_3$), 250 mg of 2-hexadecanediol (HDD), and 50 mL of diphenyl ether were mixed in a 200 mL three-neck flask and sonicated for 5 min. Next, 50 mg of commercial Pt/C (TKK, 20 wt. % Pt) dispersed in 5 mL of hexane was injected into the prepared solution. The mixture was heated to 259 °C for 30 min with stirring of 100 rpm. After the reaction, PtMn/C catalysts were synthesized and obtained from the resultant mixture solution by adding acetone and washed three times with a 40 mL of mixture of acetone and hexane (3:1). After the washing procedure, the PtMn/C catalysts were dried in a vacuum for 12 h. Atomically ordered Pt$_3$Mn/C catalysts were obtained by applying heat-treatment on the synthesized PtMn/C catalysts, at 700 °C under the reductive atmosphere (4 % H$_2$ and 96 % Ar) for 4 h.

## Preparation of atom probe specimens

In order to embed the carbon-supported nanoparticles in a metal film, we performed co-deposition of the nano-catalysts with the Co film. Co was selected as a matrix material since Co has a single isotope in nature. Co electrodeposition bath was prepared by dissolving 48 g of $CoSO_4·7H_2O$, 12 g of $CoCl_2·6H_2O$ and 9 g of $H_3BO_3$ in 250 mL of distilled (DI) water. Dried nano-catalysts were weighed and dispersed in the Co bath solution in 0.1 g $L^{-1}$ ratio. After the mixture was sonicated for 30 min at room temperature, 1 mL of colloidal solution was transferred immediately after sonication, to the vertical cell comprising of the Cu foil (0.2 $cm^2$ surface area) as a cathode and the Pt wire as an anode. Pulsed electrodeposition with 10 % duty cycle and 0.1 Hz pulse frequency was done to enhance the number density of the embedded nanoparticles and to reduce the formation of $H_2$ bubbles (Jun et al., 2021). Each specimen was electrodeposited for 8000 s.

In order to facilitate the site-specific lift-out and increase the chance of capturing the nanoparticles within the topmost region of the resulting atom probe specimen, a new approach using cross-sectional sampling was developed (see Scheme 1). In previous sample preparation method (upper part in Scheme 1), electrodeposited composite film of nano-catalysts and matrix on substrate was directly moved to dual-beam focused ion beam (FIB, Helios NanoLab 450 F1) chamber. By contrast, the composite film was vertically mounted in a conductive resin, and polished to reveal the cross-section of electrodeposited matrix in the modified method (lower part in Scheme 1). A wedge of composite material is extracted using standard lift-out procedure, and fabricated into an APT specimen by annular milling as described in the previous report (Thompson et al., 2007). We described the detailed observation of APT specimens in Supplementary information.

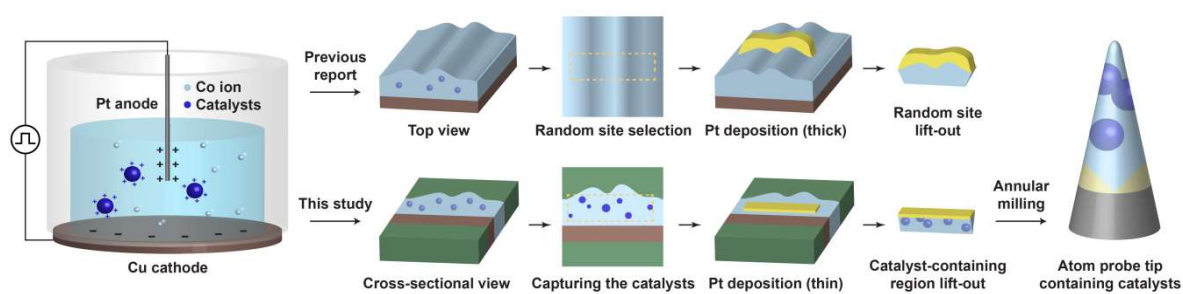

**Scheme 1.** Illustration of site-specific APT specimen fabrication procedure.

## Characterization

The crystal structures of the synthesized PtMn/C, Pt$_3$Mn/C and commercial Pt/C (TKK, 20 wt. % Pt) catalysts were determined by X-ray powder diffraction (XRD, RIGAKU SmartLab operated at 45 kV and 200 mA using Cu-Kα radiation). Bright field-transmission electron microscopy (BF-TEM) images were obtained using Tecnai TF30 ST operated at 300 kV. High-angle annular dark field scanning transmission electron microscopy (HAADF-STEM) associated with energy dispersive X-ray spectroscopy (EDX) and spherical aberration correction were performed at 300 kV (FEI Titan cubed G2 60-300). APT analyses were carried out using a local electrode atom probe (CAMECA Instruments LEAP 4000X HR$^{TM}$) at the base temperature of 50 K. The laser pulse energy and frequency were 50 pJ and 125 kHz, respectively. Data reconstruction and analyses were done with the IVAS 3.8.2 software provided by CAMECA Instruments.

## Simulations of field evaporation

On purpose of investigating the evaporation field induced artifact during the atom probe measurement, the computational simulation using the TAPSim software, developed by Oberdorfer et al. (Oberdorfer et al., 2013) was performed. Virtual APT specimens consisting of Co, with radius of 6 nm, shank angle of 20 degree, total length of 15 nm and lattice parameter of 0.36 nm was constructed using the custom Python 3.8 script. The each carbon-supported nanoparticle (Pt/C, PtMn/C, and Pt$_3$Mn/C) was included in the virtual APT specimens, respectively, based on the detected number of atoms from APT results. PtMn nanoparticle was set to have Pt core with radius of 1.320 nm and Mn shell with thickness of 0.145 nm, yielding 3:1 ratio of Pt and Mn. Pt$_3$Mn nanoparticle was set to have homogeneous ordered L1$_2$ structure. Additionally, carbon sphere with radius of 15 nm adjacent to the metallic nanoparticle was included in the specimen as the representation of carbon support. The center of the carbon sphere had the same height as the metallic nanoparticle.

Evaporation field of each metallic element in the nanoparticles were set at 37, 30, and 45 V nm$^{-1}$ each for Co, Mn and Pt, respectively, referring the theoretically calculated evaporation field values (Gault et al., 2012; Larson et al., 2013). In addition to the predicted value of 103 V nm$^{-1}$, the evaporation field of a carbon support was varied at 50, 75 and 103 V nm$^{-1}$ for systematic study. Resulting detector hit map data from TAPSim evaporation simulation was reconstructed using standard shank angle reconstruction protocol (Larson et al., 2013), and re-visualized using IVAS 3.8.2 software.

# Results and discussions

### Characterization of carbon-supported nanoparticles

Figure 1 shows XRD patterns of the Pt/C, PtMn/C, and Pt$_3$Mn/C. The background peak at 2θ of 20 to 30 degree representing amorphous carbon (Ungar et al., 2002) is observed in all samples. In both Pt/C and PtMn/C samples, the relatively board peaks are originated from the disordered FCC Pt. No Mn-related peak is found. After the reductive heat-treatment of Pt$_3$Mn/C, the ordered FCC peaks were observed indicating Mn ions were first reduced during the reaction in the solution and heat-treatment induced the chemical ordering of Pt$_3$Mn.

The average diameters of Pt/C, PtMn/C, and Pt$_3$Mn/C were measured to be 2.6 ± 0.4, 3.3 ± 0.5, and 4.2 ± 0.8 nm, respectively, (see Figure S2a-c), implying the mild agglomeration occurred during the heat-treatment at 700 °C, which is in accordance with the calculation result using the Scherrer equation (see the Supplementary information). In addition, high-resolution STEM-EDX result of the Pt$_3$Mn/C showed fully ordered L1$_2$ structure from (see Figure S2d) (Jung et al., 2017).

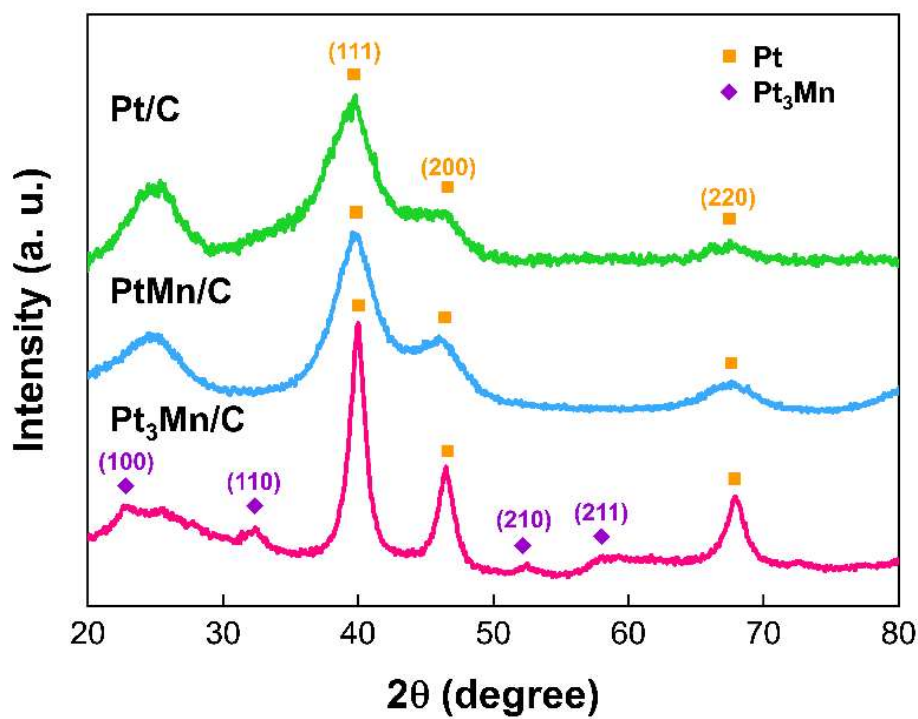

**Figure 1.** XRD patterns of Pt/C, PtMn/C, and Pt$_3$Mn/C (Pt standard: PDF no. 01-087-0647, Pt$_3$Mn standard: PDF no. 03-065-3260, and Mn$_2$O$_3$ standard: PDF no. 01-078-0390)

## APT results of carbon-supported Pt nanoparticles

Figure 2a shows the 3D atom map of the Pt/C embedded in the Co matrix. There are low-density volumes with small amount of Co atoms at the top-right and bottom-left side of the specimen (see Figure 2a). In addition, Pt and C atoms were detected with high density at the border between low density and normal region. A two-dimensional (2D) contour plot, a graphical technique for representing an atom-density surface of overall collected atoms by plotting along the measurement direction, illustrates a clear low-density region in the reconstructed specimen (see Figure 2b and c). Figure 2d and 2e show the one-dimensional (1D) concentration and detected atom profiles across the Pt particles located on the interface between nanoparticle and matrix. Measured diameters of Pt particles were 2-3 nm based on a full width half maximum of the Pt concentration in 1D concentration profile (Figure 2e). More importantly, peaks for detected atoms were located at the border of low density and normal density region implying strong LME.

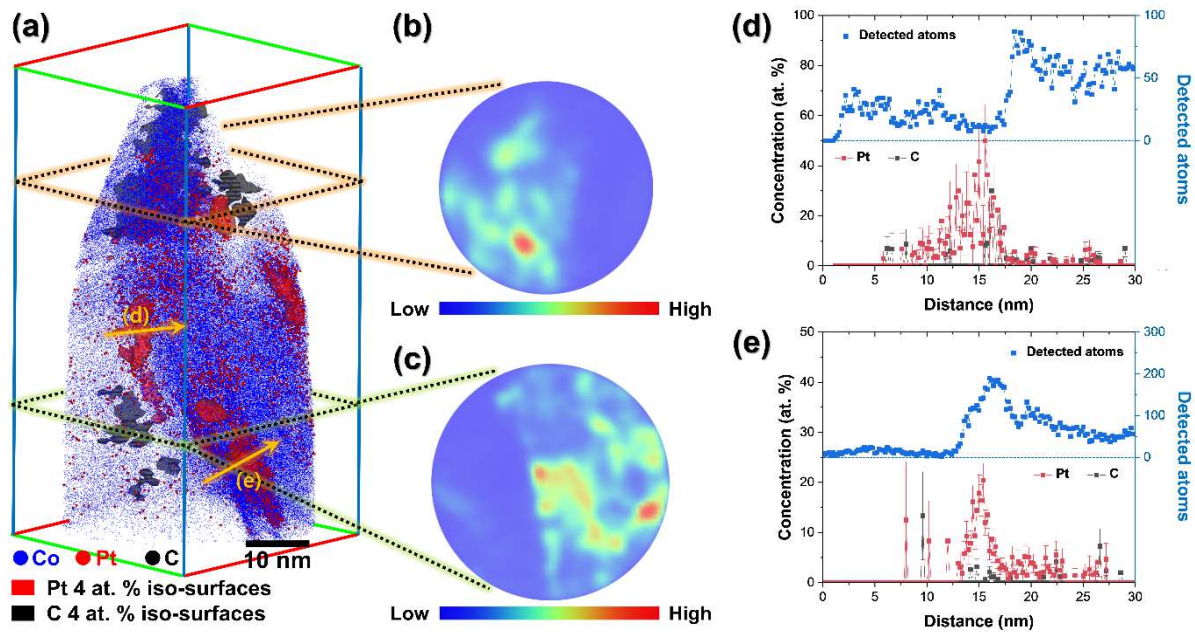

**Figure 2.** (a) 3D atom map of Pt/C and 2D atomic contour plots from (b) top and (c) bottom-side of the specimen. (d,e) 1D concentration profiles across the Pt-rich regions (direction: orange arrows in atom maps)

## Simulated results of carbon-supported Pt nanoparticles

In APT analysis data of carbon-supported metallic nanoparticles, two major artifacts, namely formation of low-density area and overlapping of consisting elements were observed. Both artifacts are reportedly to occur in atom probe analysis of specimen containing phases with large evaporation-field difference (Jun et al., 2021; Se-Ho Kim et al., 2020). To elucidate what causes the severe distortion, TAPSim analyses were conducted on each nanoparticle/carbon case. Figure 3 shows evaporation simulation result of the virtual APT specimen containing the Pt nanoparticle (red) on the carbon support (purple), performed with various evaporation field of carbon. At initial state before evaporation (Figure 3a), atomic density is uniform across the whole specimen. The 2D contour plot (Figure 3e) thus shows uniform distribution inside the nanoparticle. 1D concentration profile across the nanoparticle (Figure 3i) also shows uniform number density (marked as blue squares), regardless of whether C (marked as black squares) or Pt (marked as red squares) is enriched at the position.

As the evaporation-field value of carbon increases from 50 to 103 V $nm^{-1}$, larger distortion in shape and density of the reconstructed atom maps (Figure 3b-d) is observed. While phase with lower evaporation field (Co matrix, Pt nanoparticle) laterally contracts, phase with higher evaporation field (C support) expands, resulting in increased atomic density at Co / Pt region and severely decreased atomic density at carbon support region. 2D contour plot from each atom map (Figure 3f-h) supports this observation, showing large atomic density difference between Co / Pt region and carbon support region. In the 1D concentration profile across Pt clusters in each atom map (Figure 3j-l), clear separation of low density region and normal density region is observed. The increase of the density rises more sharply as the evaporation field of C support increases. Additionally, the peak of Pt concentration is located at the border of low density and normal density region, which is in good qualitative agreement with the APT analysis result (Figure 2). The result indicates that the large evaporation field difference between metal and carbon atom could cause the severe density fluctuation within the atom map after APT analysis of carbon-supported metallic nanoparticles.

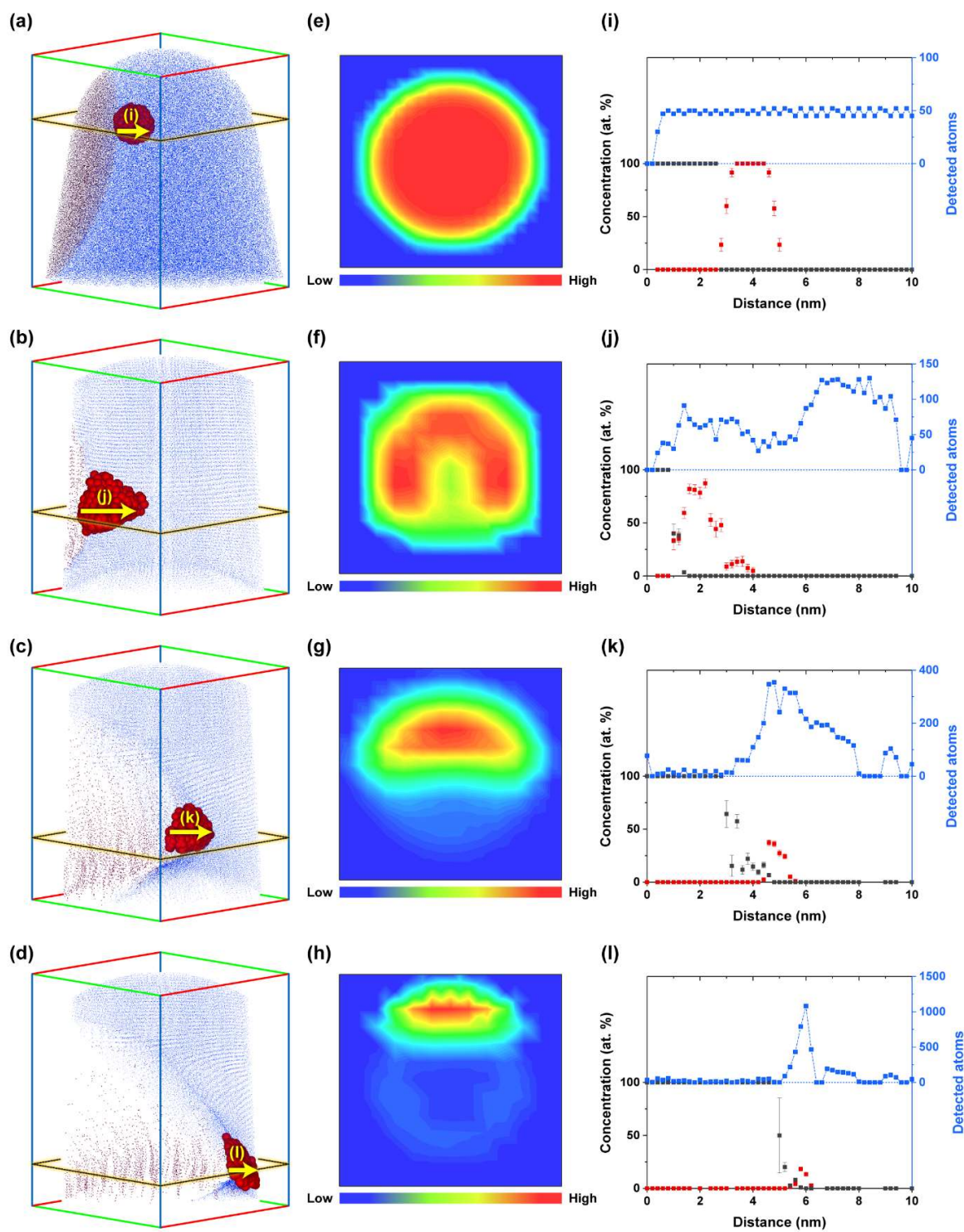

**Figure 3.** Simulated 3D atom maps of APT specimen containing Pt/C nanoparticle (a) before evaporation, after evaporation simulation of Pt/C with evaporation field of carbon set at (b) 50, (c) 75, and (d) 103 V nm$^{-1}$, (e-h) 2D contour plots and (i-l) 1D concentration profiles from each atom map (direction: yellow arrow in atom maps).

## APT results of carbon-supported PtMn and Pt$_3$Mn nanoparticles

In order to reveal the difference in the distribution of Mn atoms and the concentration of the as-synthesized PtMn/C and Pt$_3$Mn/C catalysts, APT measurements were performed. Figure 4a and 4b show 3D atom maps of the PtMn/C and the Pt$_3$Mn/C cylindrically cut with diameter of 25 nm and thickness of 25 nm, including iso-concentration surfaces of 2.5 at. % C (black surfaces) and 4 at. % Pt (red surfaces). The high evaporation-field difference among multi-elements (Pt, Mn, Co and C) as well as uneven evaporation of carbon complexes result in overlapping of the elements between nanoparticles and matrix. Pt-rich volumes represent the PtMn or Pt$_3$Mn nanoparticles. The resulting compositions of nanoparticles were 73.8 ± 1.2 at. % of Pt and 26.2 ± 1.2 at. % of Mn for PtMn nanoparticle and 79.8 ± 0.8 at. % of Pt and 20.2 ± 0.8 at. % of Mn for Pt$_3$Mn nanoparticles, respectively, which are close to the composition of Pt$_3$Mn intermetallic compound.

Figure 4c and 4d show detected Pt and Mn atoms in the nanoparticles. The majority of Mn atoms were measured at the center of the nanoparticles in Pt$_3$Mn/C while in PtMn/C, the atoms were observed near the surfaces. In addition, the concentration profile of PtMn/C shows Mn enrichment at the outer of the particle indicating typical core-shell structure while Mn composition was high at the inside of the particle implying typical homogeneous distributed concentration profile, as shown in Figure 4e and 4f.

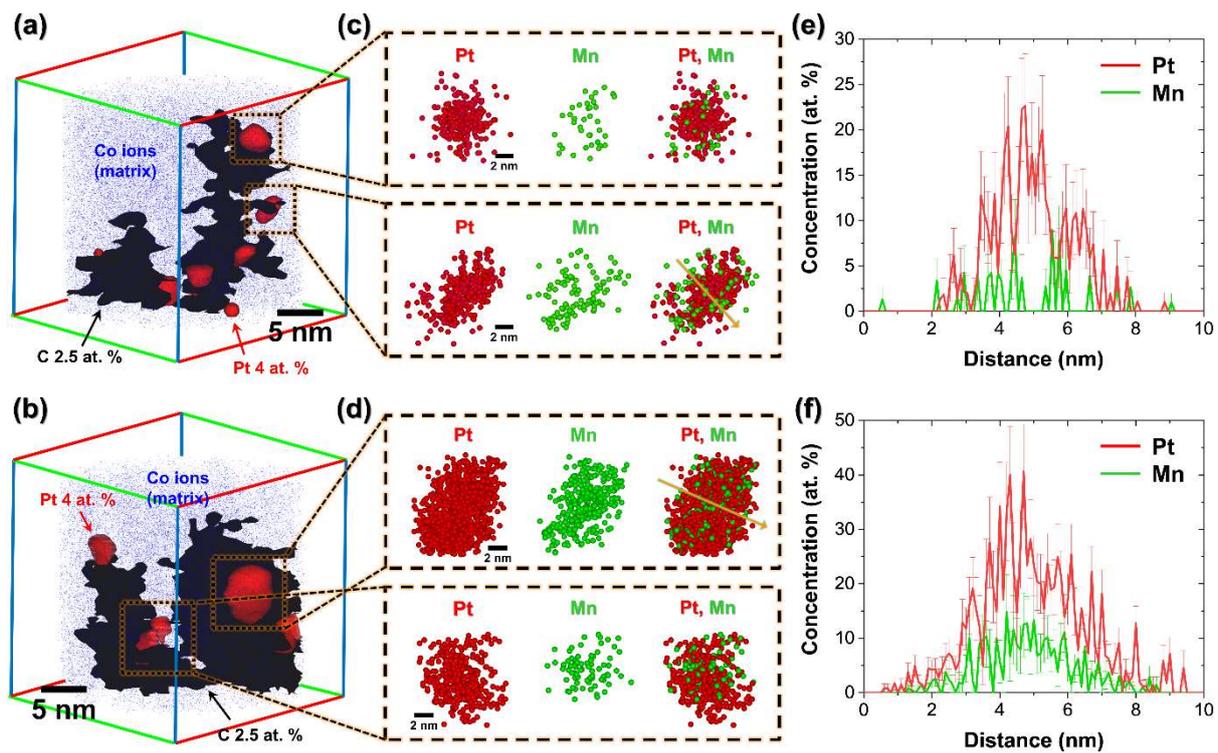

**Figure 4.** 3D atom maps of (a) PtMn/C and (b) Pt$_3$Mn/C, including iso-concentration surfaces of 2.5 at.% C and 4 at.% Pt. Sectioned maps displaying Pt (red) and Mn (green) atoms and corresponding 1D concentration profiles along the orange arrows in (c,e) PtMn/C and (d,f) Pt$_3$Mn/C nanoparticle.

# Simulated results of carbon-supported PtMn nanoparticles

Since the computational simulation has shown that position and concentration artifact could take place in near region to the carbon support, the reliability of the relative position of Mn and Pt atoms could be also questioned. In order to find out whether the distribution of Mn atoms within nanoparticles could be analyzed under the effect of carbon support, evaporation simulation with APT specimens containing PtMn and $Pt_3Mn$ nanoparticles with carbon support were performed, respectively. Figure 5a-d shows the 3D atom maps of specimen containing PtMn nanoparticle before evaporation (Figure 5a), and reconstructed after evaporation simulation with various evaporation field of carbon (Figure 5b-d). Pt and Mn atoms are shown in red and green spheres in each atom maps. 1D concentration profiles taken in direction of yellow arrow are presented in Figure 5e-h.

At initial state before evaporation, the core of the nanoparticle is enriched with Pt (red spheres), and Mn atoms (green spheres) are distributed on the surface of the nanoparticle as shown in Figure 5a. As a result, 1D concentration profile across the nanoparticle (Figure 5e) shows a single concentration peak of Pt, with two separated peaks of Mn at each end of the Pt concentration peak. After evaporation simulation, distortions resulted from LME are observed in the reconstruction maps (Figure 5b-d). Although the discrepancy from the initial state gets larger as the field of C increases, the concentration profile across each nanoparticle system (Figure 5f-h) shows the Mn atoms are enriched at the surfaces of the particle.

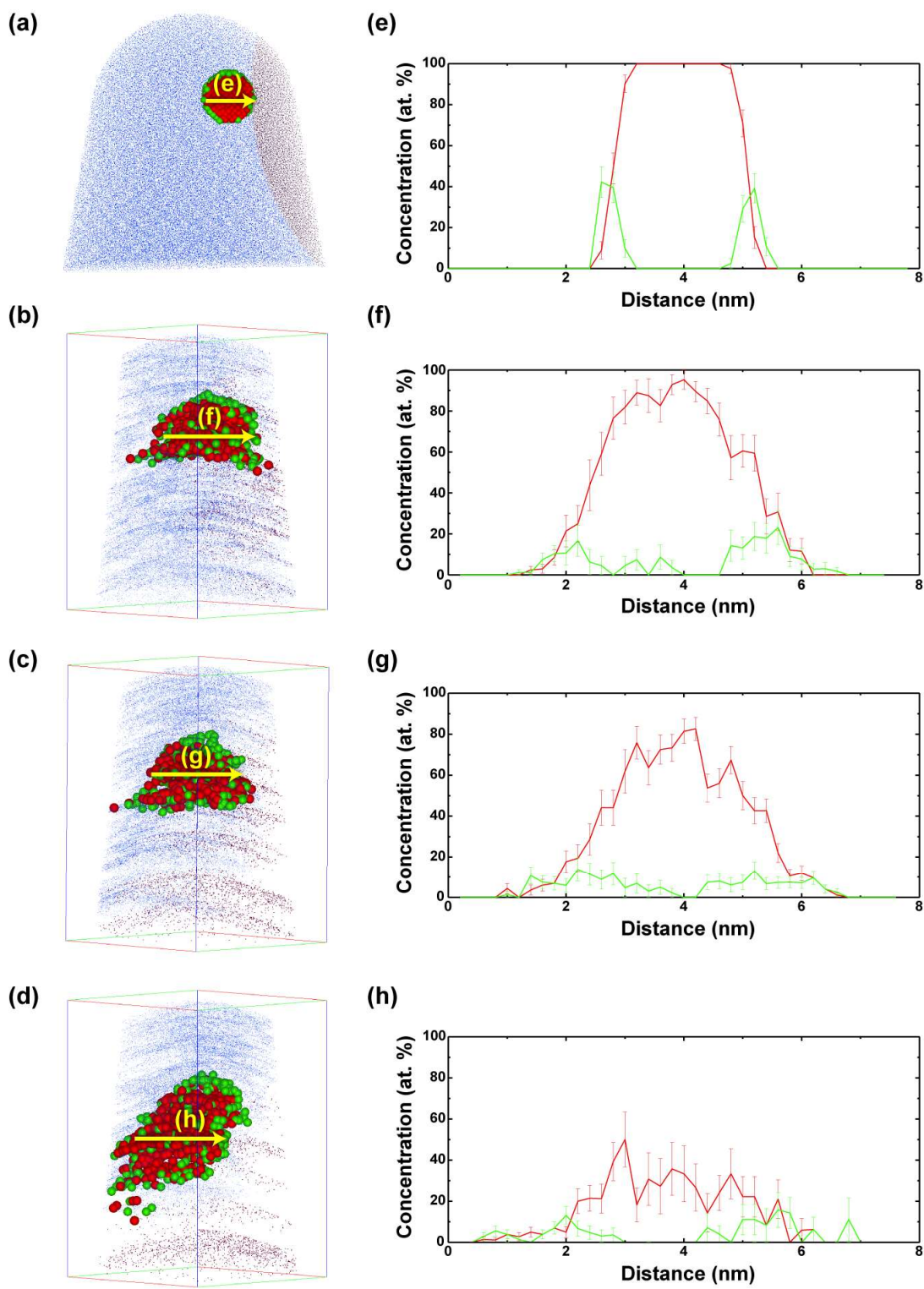

**Figure 5.** Simulated 3D atom maps of PtMn/C (a) before evaporation and after evaporation with evaporation field of Carbon set at (b) 50 V, (c) 75 V, and (d) 103 V nm$^{-1}$, and (e-h) 1D concentration profiles of Pt, Mn from each atom maps (direction: yellow arrow in atom maps).

**Simulated results of carbon-supported Pt$_3$Mn nanoparticles**

Figure 6a illustrate the initial simulation condition of the ordered L1$_2$ Pt$_3$Mn particle on carbon support embedded in the Co matrix. The same protocol is used as above and the reconstructions of the particles are severely distorted and exhibited a compressed shape (Figure 6b-d). Unlike the PtMn nanoparticles results in which the Mn atoms are only enriched at the surface, Mn are distributed homogeneously with 3:1 Pt/Mn atomic ratio in all Pt$_3$Mn nanoparticles (see Figure 6f-h). A clear difference between PtMn and Pt$_3$Mn results was observed in both atom maps and 1D concentration profiles even after applying extremely high evaporation field of the carbon support. Field evaporation of carbon indeed influence the acquired atom map however, these results suggest that qualifying and quantifying of 3D elemental distributions of carbon supported nanoparticles are possible from APT reconstruction map.

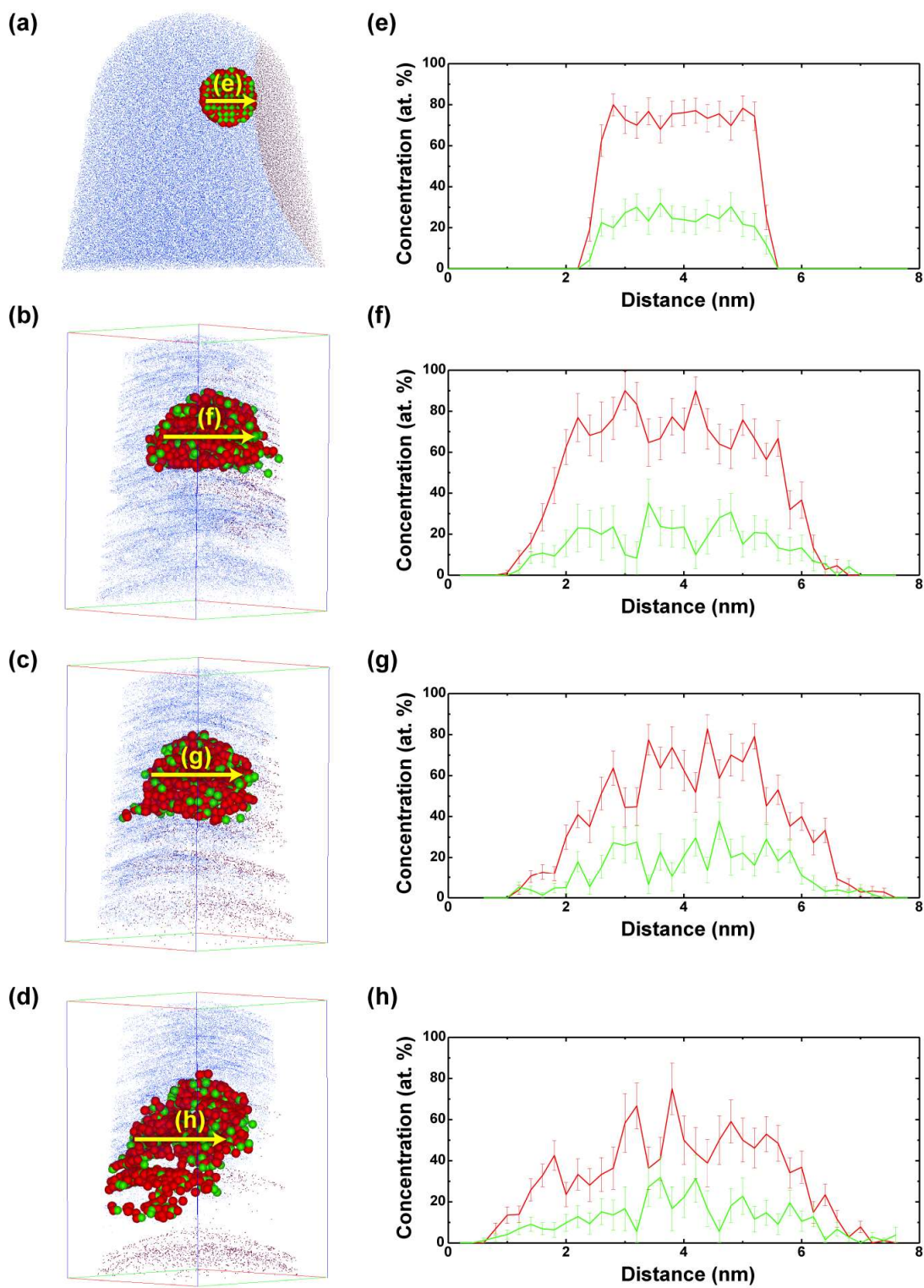

**Figure 6.** Simulated 3D atom maps of $Pt_3Mn/C$ (a) before evaporation and after evaporation with evaporation field of Carbon set at (b) 50, (c) 75, and (d) 103 V $nm^{-1}$, and (e-h) 1D concentration profiles of Pt, Mn from each atom maps (direction: yellow arrow in atom maps).

## Mn diffusion in the nanoparticles during the synthetic process

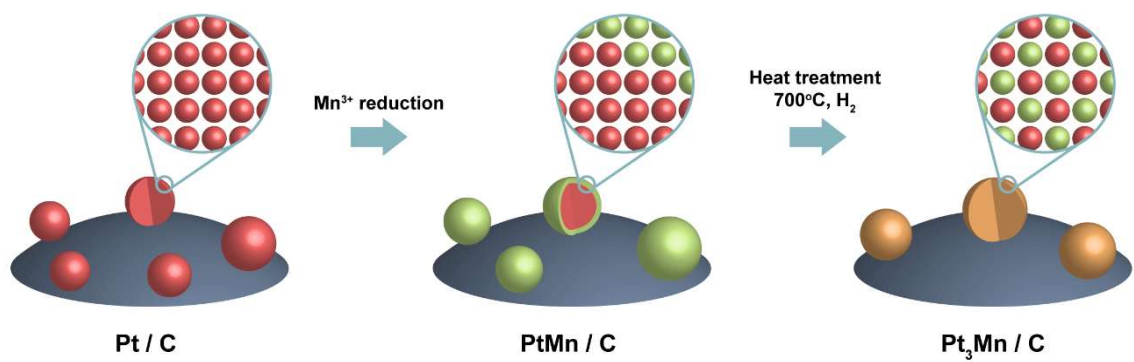

**Scheme 2**. Fabrication process of atomically ordered Pt$_3$Mn/C

Synthesis of atomically ordered Pt$_3$Mn/C requires two-step process: reduction of Mn ions in solution and heat-treatment in reductive atmosphere (see Scheme 2). We observe the Mn enrichment at the surface of the particles in PtMn/C and homogeneous 3:1 Pt/Mn distribution in Pt$_3$Mn/C from APT experiment (see Figure 4). In addition, simulated results of carbon-supported PtMn and Pt$_3$Mn nanoparticles support the different tendency (see Figure 5 and 6). Thus, we suggest that Mn ions were reduced and only attached to the surface of Pt nanoparticles after chemical reaction in solution. Subsequently, during the reductive heat-treatment, surface-reduced Mn atoms diffused into the core of the particles and resulted in the ordered L1$_2$ structure. Chi et al. reported the surface segregation of Pt in Pt$_3$Co nanoparticle starts at 350 ºC while L1$_2$ ordering in Pt$_3$Co nanoparticles appears at 600 ºC (Chi et al., 2015), implying the applied heat (259 ºC) may not provide enough thermal driving force to initiate diffusion toward the particle core. Another possible reason could be Mn ions were not fully reduced instead binding with oxygen at the particle surface that may hinder the movement of Mn atoms. The Mn$_x$O$_y$ molecular ion peaks were detected in the acquired mass spectra and the intensity in PtMn/C was higher than that of Pt$_3$Mn/C (see Figure S3).

# Conclusions

In this study, we reveal the 3D elemental distribution of carbon supported Pt, PtMn and Pt$_3$Mn nanoparticles. Systematic numerical simulations with varied evaporation-field conditions of the carbon are used to study the LME. It has shown that despite the artifacts in atomic density and the shape, the atomic movement within the nanoparticle during the heat-treatment could have been tracked with APT. We envision that the presented study can pave the road for the application of APT analysis for various carbon supported nanomaterials.

# Acknowledgments

This research was supported by the National Research Foundation of Korea (NRF) (grant numbers 2020R1A6A3A13073143 and 2021R1A6A3A03045488). S.-H.K. acknowledges the financial support from the ERC-CoG-SHINE-771602. The authors thank Dr. Woojin Yun from National NanoFab Center (NNFC) and Mr. Tae-Woo Lee from KAIST Analysis center for Research Advancement (KARA) for supporting the experiments.

# Supplementary Information

# Tracking the Mn diffusion in the carbon-supported nanoparticles through the collaborative analysis of atom probe and evaporation simulation

Chanwon Jung [a,b,†], Hosun Jun [a,†], Kyuseon Jang [a], Se-Ho Kim [b], Pyuck-Pa Choi [a,*]

[a] Department of Materials Science and Engineering, Korea Advanced Institute of Science and Technology (KAIST), 291 Daehak-ro, Yuseong-gu, Daejeon 34141, Republic of Korea

[b] Max-Planck-Institut für Eisenforschung GmbH, Düsseldorf 40237, Germany

[†] These authors contributed equally to this work

[*] Corresponding Author: p.choi@kaist.ac.kr

**Key words**

Atom probe tomography, local magnification effect, carbon-supported nanoparticles, field evaporation simulation, intermetallic compounds

## Fabrication of APT specimen including carbon-supported nanoparticles

Figure S1a shows SEM image of the electrodeposited composite film of the nano-catalysts and the Co matrix. As backscattered electron (BSE) mode has a moderately high sensitivity in terms of atomic number, distinguished brightness indicates the different materials of the carbon containing nano-catalysts (dark area) and the matrix (bright area). Co matrix containing a relatively high density of the nano-catalysts (Figure S1b) is selected and sliced using FIB for lift-out. Figure S1c shows a composite lamella before lift-out process with the completely embedded nano-catalysts. After annular milling, sharpened APT specimen containing nano-catalysts was obtained, as shown in Figure S1d.

The final APT specimens were measured with bright-field TEM to first confirm possible existence of pores that result abruptly fracture during high-field evaporation process. No noticeable crack or voids were observed in the specimen (see Figure S1e). In bright-field mode, elements with lower atomic mass show brighter contrast than elements with higher atomic mass. Thus, bright features showing sphere shape are the supporting carbon, while darker region is the Co matrix. Figure S1f shows HAADF-STEM image of sharpened APT specimen showing different contrast. In addition, even brighter spheres in size of few nanometers representing PtMn nanoparticles were observed without shape change, indicating the electrodeposition process has successfully captured the PtMn nanoparticles within matrix, without affecting the morphology of nano-catalysts (see Figure S1g).

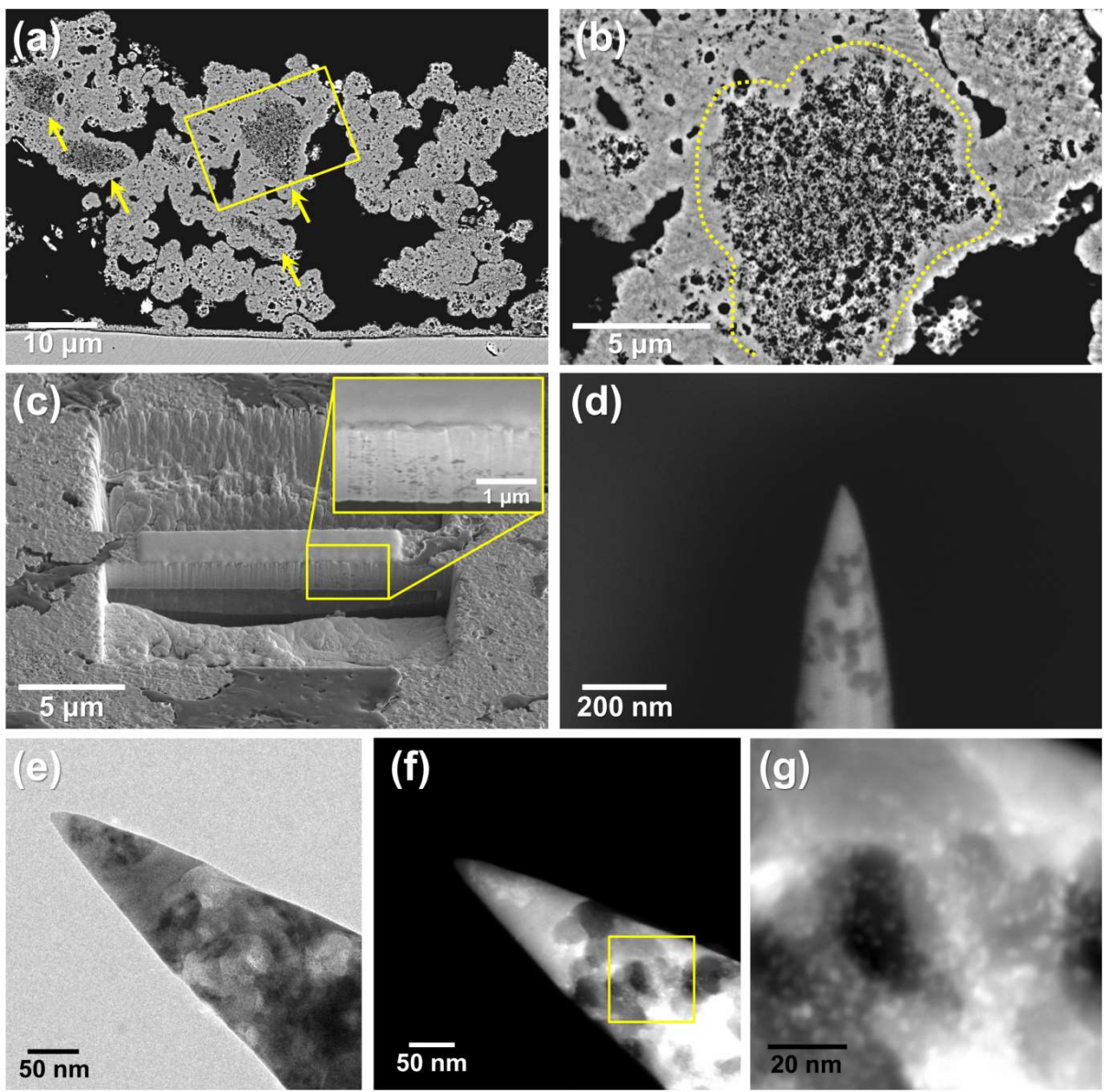

**Figure S1.** SEM images of (a) electrodeposited composite film of nano-catalysts and matrix (yellow arrows indicating embedded nano-catalysts and yellow box representing the area in b), (b) close-up image of region of interest (dotted line showing the nano-catalysts), (c) cross-sectional view after lift-out process and (d) sharpened APT tip. (e) BF-TEM and (f-g) STEM images of prepared APT tip.

# Calculation of average particle size from XRD results

The 2-theta peaks from FCC (111) were centered at 39.71°, 39.77°, and 40.00° for Pt/C, PtMn/C, and Pt$_3$Mn/C, respectively. No shift between Pt/C and PtMn/C indicates that enough mixing between Pt and Mn did not occur during the synthesis of PtMn/C. By contrast, 2-theta of Pt$_3$Mn/C shows larger than PtMn/C indicating a reduction of the lattice constant. The average particle size was estimated by using the Scherrer equation as below (Scherrer, 1918).

$$D_{hkl} = \frac{K \lambda}{B_{hkl} \cos\theta}$$

Here, $D_{hkl}$ is the average crystallite size in direction of hkl plane, K is crystallite shape factor which we used 0.9 (sphere shape) in this study (Holzwarth & Gibson, 2011), $\lambda$ is the wavelength of the X-ray, $B_{hkl}$ is the full-width at half-maxium of the X-ray diffraction peak from hkl and $\theta$ is the Bragg angle. The estimated average particle size of Pt/C, PtMn/C, and Pt$_3$Mn/C were 2.1, 2.5, and 6.7 nm, respectively.

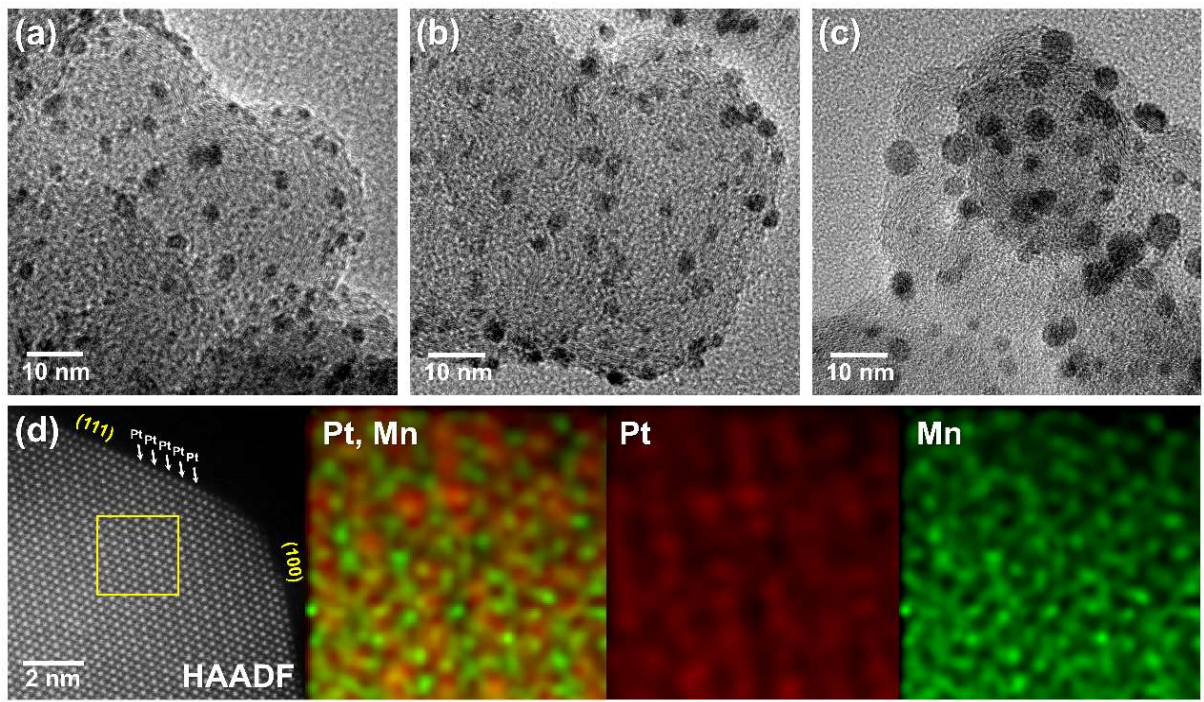

**Figure S2.** BF-TEM images of (a) Pt/C, (b) PtMn/C, and (c) Pt$_3$Mn/C. STEM-EDX maps of (d) High-resolution STEM images and EDX maps of Pt$_3$Mn/C in atomic resolution.

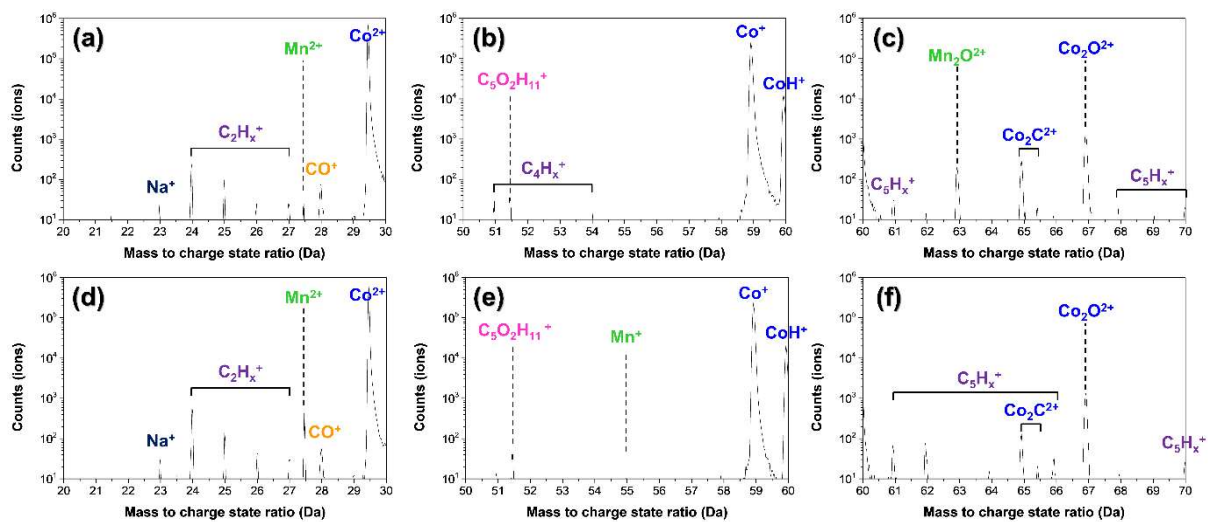

**Figure S3.** Acquired mass spectra from (a-c) PtMn/C and (d-f) Pt$_3$Mn/C for closer inspection of the Mn and Mn$_2$O ions peaks